\documentclass[11pt]{article}
\usepackage{moriond,epsfig}
\usepackage{subfigure}

\newcommand{\vect}{\vec}

\begin{document}
\vspace*{4cm}
\title{Experimental measurement of muon $(g-2)$}

\author{F.E. Gray for the Muon $(g-2)$ Collaboration~\footnote{
"R.M.~Carey, E.~Efstathiadis, M.F.~Hare, X.~Huang,
F.~Krienen, A.~Lam, I.~Logashenko, J.P.~Miller, J.~Paley, Q.~Peng, O.~Rind,
B.L.~Roberts, L.R.~Sulak, and A.~Trofimov (Boston University);
G.W.~Bennett, H.N.~Brown, G.~Bunce, G.T.~Danby, R.~Larsen, Y.Y.~Lee, W.~Meng,
J.~Mi, W.M.~Morse, D.~Nikas, C.~{\"O}zben, R.~Prigl,
Y.K.~Semertzidis, and D.~Warburton (Brookhaven National Laboratory);
Y.~Orlov (Cornell University);
A.~Grossmann, G.~zu~Putlitz, and P.~von~Walter (Universit{\"{a}}t Heidelberg);
P.T.~Debevec, W.~Deninger, F.E.~Gray, D.W.~Hertzog,
C.J.G.~Onderwater, C.~Polly, M.~Sossong, and D.~Urner (University of Illinois at
 Urbana-Champaign);
A.~Yamamoto (KEK);
K.~Jungmann (Kernfysisch Versneller Instituut);
B.~Bousquet, P.~Cushman, L.~Duong,
S.~Giron, J.~Kindem, I.~Kronkvist, R.~McNabb, T.~Qian, and P.~Shagin (University
 of Minnesota);
V.P.~Druzhinin, G.V.~Fedotovich, D.~Grigoriev,
B.I.~Khazin, N.M.~Ryskulov, Yu.M.~Shatunov, and E.~Solodov (Budker Institute of 
Nuclear Physics);
M.~Iwasaki (Tokyo Institute of Technology);
M.~Deile, H.~Deng, S.K.~Dhawan, F.J.M.~Farley,
V.W.~Hughes~(deceased), D.~Kawall, M.~Grosse-Perdekamp, J.~Pretz, S.I.~Redin,
E.~Sichtermann, and A.~Steinmetz (Yale University).}}

\address{Dept. of Physics, U. of Illinois at Urbana-Champaign,
1110 W. Green St., Urbana, IL  61801, U.S.A.
\footnote{Present address: Dept. of Physics, U. of California, Berkeley, 366 LeConte Hall, Berkeley, CA  94720, U.S.A.}
}

\maketitle\abstracts{The muon $(g-2)$ experiment at Brookhaven National
Laboratory has measured the anomalous magnetic moment of the positive
muon with a precision of 0.7~ppm.  This paper presents that result, 
concentrating on some of the important experimental issues that arise in
extracting the anomalous precession frequency from the data.}

\section{Concept}

\begin{figure}[tb]
\begin{center}
\subfigure[]{\includegraphics[height=5cm]{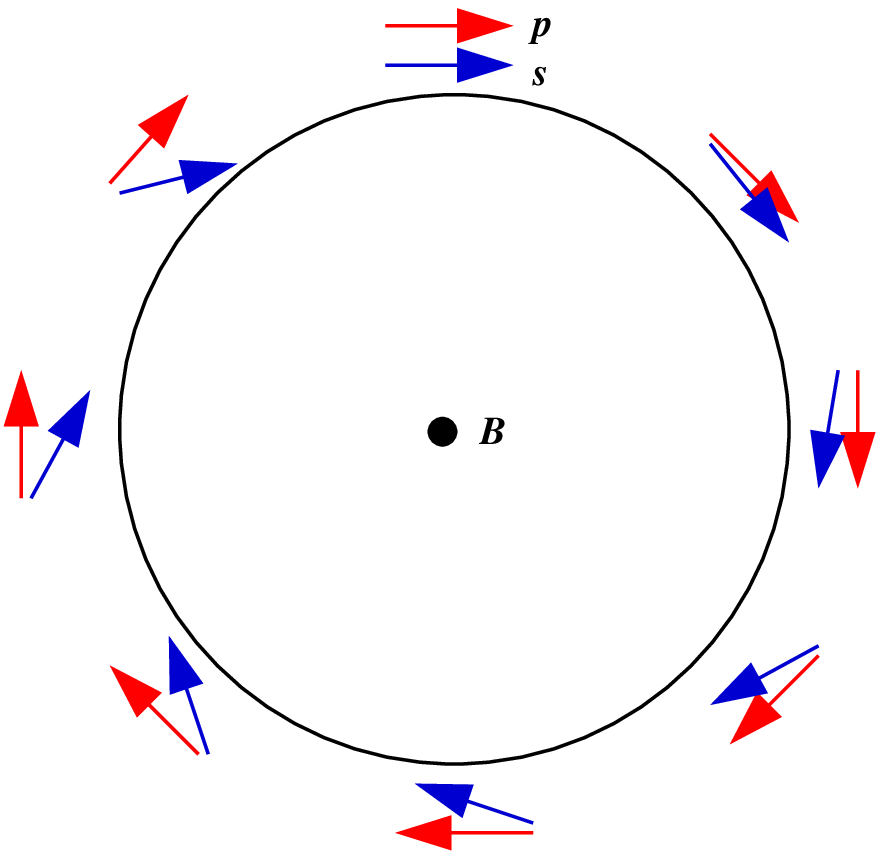} \label{concept}}
\subfigure[]{\includegraphics[height=7cm]{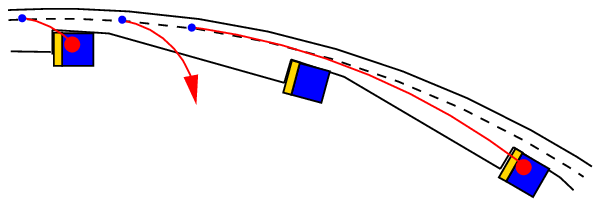} \label{decay}}
\end{center}
\caption{(a) Motion of the muon spin vector relative to its
momentum vector.  (b) Trajectories followed by several decay positrons of
varying energies.}
\label{concept-decay}
\end{figure}

The spin of the muon generates a magnetic moment whose strength is described
by a dimensionless quantity $g_\mu$, the gyromagnetic ratio:
\begin{displaymath}
\vect{\mu} = g_\mu \left(\frac{e}{2m}\right) \vect{s} ~.
\end{displaymath}
The Brookhaven National Laboratory AGS E821 muon $(g-2)$ experiment
measures the anomalous part $a_\mu$ of the gyromagnetic ratio, defined by 
$a_\mu = \frac{1}{2}(g_\mu - 2)$.
For a pointlike Dirac particle, $a_\mu = 0$.  It acquires a nonzero value
only through radiative corrections.  It may be measured experimentally with
great precision; the published data from E821 have a precision of 0.7 parts 
per million (ppm).  It may also be calculated precisely in the context of the 
standard model, taking into account contributions from the electromagnetic and 
weak interactions, which are understood well, and from hadronic processes, 
which are more troublesome.  The current state of these calculations is 
described in this volume by A.~Nyffeler and 
A.~H\"ocker.  The nominal precision of these results is also at the level
of 0.7~ppm, but inconsistent values are obtained depending on the hadron 
production cross section data that constitute an essential input.  
Once this situation is resolved, it will be possible 
to draw conclusions from a comparison of the experimental and theoretical
results.  Such a comparison may either yield evidence for or a constraint 
against physics beyond the standard model.  

The idealized muon $(g-2)$ experiment places a polarized ensemble of
muons in a uniform magnetic field at $t=0$.  They follow circular orbits
in this field with a cyclotron frequency of 
\begin{displaymath}
\vect{\omega}_c = \frac{e \vect{B}}{m \gamma} ~.
\end{displaymath}
Meanwhile, the field rotates their spin at a frequency of 
\begin{displaymath}
\vect{\omega}_s  = g_\mu \frac{e}{2m} \vect{B} +
   (\gamma - 1) \frac{e \vect{B}}{m \gamma} ~.
\end{displaymath}
This frequency includes a term resulting from the Thomas 
precession~\cite{Thomas:1927} because the muon is in a rotating
reference frame.  The difference between the cyclotron and spin precession 
frequencies is the so-called anomalous precession frequency:
\begin{equation}
\vect{\omega}_a = \vect{\omega}_s - \vect{\omega}_c =
   \frac{e}{m} a_\mu \vect{B} ~.
\label{omegaA}
\end{equation}
This frequency is the rate at which the spin turns with respect to the
momentum, and it is therefore the apparent precession frequency from the point 
of view of an observer in the laboratory, as shown in Figure~\ref{concept}.
It is proportional to $a_\mu$, not to $g_\mu$, so $a_\mu$ may in principle
be determined directly by measuring $\omega_a$ and $B$ and performing a little 
arithmetic.  In practice, though, the magnetic field is mapped using 
NMR magnetometers,\cite{nmr} which measure the spin precession frequency 
$\omega_p$ of protons at rest in the field.  To minimize the number of
external constants required, Equation~\ref{omegaA} is rewritten as
\begin{eqnarray*}
a_\mu & = & \frac{\frac{\omega_a}{\omega_p}}{\frac{\omega_s}{\omega_p}-\frac{\omega_
a}{\omega_p}}  = \frac{R_\omega}{\lambda - R_\omega} ~{\rm with} \\
R_\omega  &=& \frac{\omega_a}{\omega_p} ~{\rm and}~ 
\lambda = \frac{\omega_s}{\omega_p} = \frac{\mu_\mu}{\mu_p} = 3.183\,345\,39(10) ~.
\end{eqnarray*}
This value of $\lambda$ is determined from experiments on the 
hyperfine structure of muonium conducted at Los Alamos National 
Laboratory, together with some theoretical input.\cite{Liu:1999iz,Nio:1997fg}  

Eventually, essentially all of the muons decay via the three-body 
process $\mu^+\rightarrow~e^+\nu_e\bar\nu_\mu$.  Because the resulting 
positrons have a lower energy than that of the muon, they curl in toward
a detector as shown in Figure~\ref{decay}.  In the CM frame, the kinematic 
distribution of the decay positrons is peaked along the direction of the spin 
of the parent muon:~\cite{Farley}
\begin{displaymath}
\frac{dP}{dy~d\Omega}  =  n(y)[1+A(y) \cos \theta_s] ~{\rm where} 
\end{displaymath}
\begin{displaymath}
n(y) = y^2(3-2y) ~{\rm and}~ A(y) = \frac{2y-1}{3-2y} ~.
\end{displaymath}
In these expressions, $y = 2E/(m_\mu-m_e)$ is the normalized energy of the 
positron and $\theta_s$ is the CM angle between the muon's spin and the 
positron's momentum.  As the spin turns with a frequency $\omega_a$, 
this ``searchlight'' of decay positrons moves with it.  Approximating both the 
muon and positron as fully relativistic particles so that $E=p$, the boost 
into the laboratory frame is described by
\begin{displaymath}
E_{lab} = \gamma E_{CM} (1 + \cos\theta_{CM}) ~,
\end{displaymath}
where $\theta_{CM}$ is the CM angle between the positron momentum and the
forward direction defined by the muon momentum.  To have a high energy in the 
laboratory, a decay positron must therefore both have a high energy in the
CM frame and also be directed forward.
Consequently, the boost translates the angular sweep of the CM frame into a
modulation of the energy distribution in the laboratory frame.  By counting
the number of decay positrons exceeding a laboratory energy threshold as a
function of time after injection, one obtains a spectrum that is described by
the functional form
\begin{equation}
N(t) = N e^{-t/\tau} [1 - A \cos(\omega_a t + \phi_a)] ~.
\label{idealFivePar}
\end{equation}
The frequency $\omega_a$ may then be determined by fitting this functional 
form to the observed spectrum.

The real-world experiment is quite similar to this simple outline.
The magnetic field is provided by a C-shaped superferric
storage ring magnet~\cite{Danby:2001eh}.  Its 1.45~T field is uniform at the
level of 1~ppm over the storage region after averaging over the azimuthal
coordinate.  The muon beam enters the ring through a field-free
region produced by a superconducting inflector magnet.\cite{Yamamoto:2002bb}
From this point, a circular trajectory would simply lead it around for a single
turn where it would be lost on the inflector housing.  Therefore, it is
necessary to kick it onto the central orbit with a pulsed magnet.\cite{Kicker}
To avoid perturbing the field during the measuring period, the kicker contains
no iron, and its pulse is only a little longer than one cyclotron period.

In the radial dimension, the uniform dipole field has a focusing effect: all 
circular orbits are closed regardless of the initial radial position and angle,
provided only that they do not strike an obstruction.  In the vertical 
dimension, however, additional focusing is required: a particle with even
a rather small initial vertical angle will quickly spiral up or down without 
bound out of the storage volume.  Consequently, electric 
quadrupoles~\cite{quads} are placed inside the storage ring vacuum chambers.
They are plates on which a static electrical charge is placed during the
measuring period, leading to a linear restoring force for 
particles that are off-center vertically.   The electric field appears as an
additional magnetic field in the muon's rest frame, so 
Equation~\ref{omegaA} is modified:~\cite{Bargmann:1959,Jackson}
\begin{displaymath}
\vect{\omega}_a = \vect{\omega}_s - \vect{\omega}_c =
  \frac{e}{m} \left[ a_\mu \vect{B} -
  \left( a_\mu - \frac{1}{\gamma^2 - 1} \right) (\vect{\beta} \times \vect{E}) \right] ~.
\end{displaymath}
However, the term proportional to $(\vect{\beta} \times \vect{E})$ is
eliminated by choosing a ``magic'' $\gamma \approx 29.3$, corresponding to a 
muon momentum of $3.09~{\rm GeV}/c$.  Consequently, it is not necessary to 
have precise knowledge of the quadrupole field.

\section{Determination of energies and times}

Decay positrons are detected by lead/scintillating fiber
electromagnetic calorimeters~\cite{Sedykh:2000ex} located inside the storage 
ring.  The signals from these calorimeters are recorded by 
waveform digitizers, which sample the photomultiplier output at 400~MHz.
The resulting waveforms are similar to traces on a digital oscilloscope. 
They are processed into energies and arrival times by the analysis software.
 
A potential systematic bias arises
from overlapping pulses in the calorimeters.   The $(g-2)$ phase $\phi_a$ in 
Equation~\ref{idealFivePar} is determined primarily by the time of flight 
from the decay vertex to the detector.  Consequently, it varies with positron 
energy by about 20~mrad from 1.4~to~3.2~GeV.  Overlapping pulses appear to 
have a high energy, but they carry the phase of their lower-energy 
constituents.  The concentration of overlapping pulses varies from early to 
late times after injection because their number is proportional to the square 
of the instantaneous muon decay rate.   They cause the average phase to shift 
as a function of time, thereby pulling the measured frequency. 

The first line of defense against overlapping pulses is separation.  The
algorithm used to determine times and energies from the waveforms is capable
of resolving pulses that arrive as little as 3.5~ns apart.  The procedure
is based on the principle that the recorded samples look like an averaged
pulse shape, translated in time and scaled with energy.  First, an average
pulse shape is constructed for each detector.  Then, an optimization procedure
is applied to each digitized interval, varying the assumed time, amplitude,
and pedestal to minimize the least-squares difference
\begin{displaymath}
\sum_{i\,\in\, {\rm samples}}
  [S_{i}-P-\sum_{j\,\in\, {\rm pulses}} A_j f_i(t_j)]^2 
\end{displaymath}
between the recorded samples $S_{i}$ and the average pulse shape $f(t)$.
When the fit to a single pulse is insufficient, the model is extended to 
include additional pulses.

Nevertheless, some residual overlapping pulses inevitably remain.  They are 
subtracted by forming out-of-time coincidences.  Each pulse defines a time
region of approximately 40~ns which is guaranteed to have been digitized by
the WFD.  Additional pulses found in these regions are artificially combined
to simulate the distribution of true overlapping pulses.  These constructed
distributions may then be subtracted from the data or, equivalently,
incorporated into the fitting function.

\section{Betatron motion}

\begin{figure}[tb]
\begin{center}
\subfigure[]{\includegraphics[width=6cm]{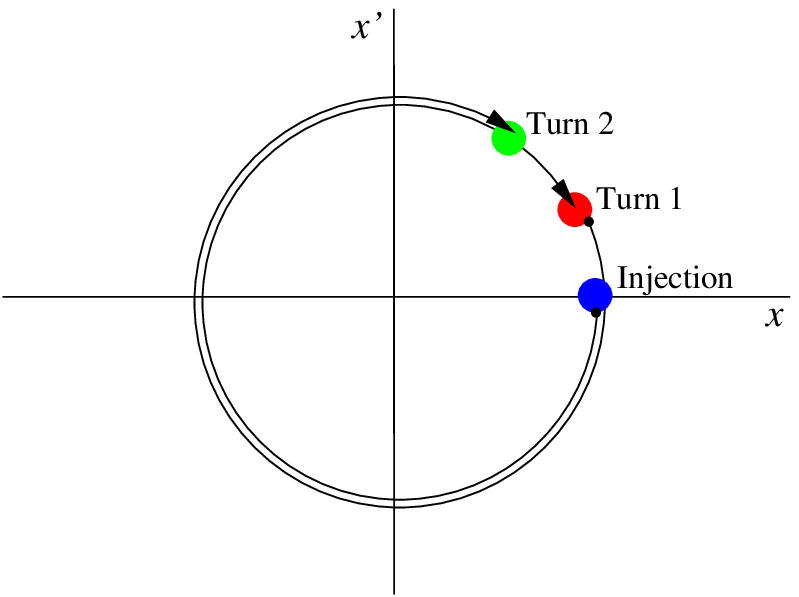}}
\subfigure[]{\includegraphics[width=6cm]{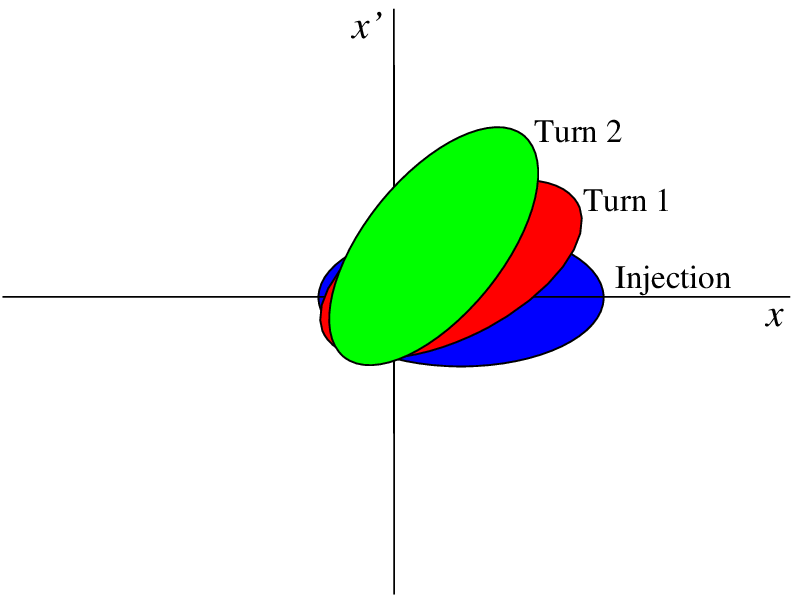}}
\end{center}
\caption{The radial betatron motion of (a) a single particle, and
(b) a distribution of particles, illustrating why the CBO appear at
the beat frequency $\omega_c - \omega_x$.  The centroid and width of the
projections of the distribution onto the $x$ and $x'$ axes clearly move with
each turn as well.}
\label{cboEllipses}
\end{figure}

The confining electric quadrupole field, together with the main dipole 
field, leads to simple harmonic motion of each particle in both the radial and 
vertical dimensions.  It is described by
\begin{eqnarray*}
x(t) & = & x_e + A_x \cos(\omega_x t + \phi_x) \\
y(t) & = & A_y \cos(\omega_y t + \phi_y) 
\end{eqnarray*}
where
\begin{displaymath}
\omega_x =  \omega_c \sqrt{1-n} ~,
\omega_y = \omega_c \sqrt{n} ~{\rm and}~ 
\end{displaymath}
\begin{displaymath}
n = \frac{\partial E_r}{\partial r} \frac{R_0}{\beta B_0} \approx 0.137 
\end{displaymath}

If the accepted phase space of the storage ring were uniformly populated, then
the betatron motion would not be relevant.  Although the individual particles
would oscillate, the $(x,x')$ and $(y,y')$ distributions as a whole would not.
However, for practical reasons, the inflector is not matched to the storage 
ring, so only a subset of the phase space is filled.  As viewed by an observer 
standing at a single azimuthal position inside the storage ring, the radial
phase space distribution appears to rotate at the beat frequency 
$\omega_{CBO} = \omega_c - \omega_x$.  The beam centroid oscillates radially
toward and away from the detector.  While the beam moves toward a detector 
on one side of the ring, it moves away from the detector on the opposite 
side.  Thus, CBO effects vary smoothly through $2 \pi$ around the storage ring.
There is also a small effect from oscillations of the width of the 
distribution.  These phenomena are illustrated in Figure~\ref{cboEllipses}.

The CBO are visible in the recorded time spectrum primarily because the detector
acceptance is a function of the decay vertex radius as well as azimuthal 
position and positron energy.  This acceptance function is established by two 
competing mechanisms.  Geometrically, particles that come from a smaller 
radius are more likely to hit the detector than those from large radii because 
their momentum has a larger azimuthal component.  However, they also pass
through obstacles such as the quadrupole or kicker plates at a more glancing
angle, increasing the probability that they will begin their electromagnetic
shower there, outside the detector.  The balance between these two mechanisms
shifts as a function of positron energy.  The fact that the acceptance is 
a function of radial position leads to an overall modulation of the time 
spectrum at the frequency $\omega_{CBO}$, at the level of 1~percent.  
Meanwhile, the fact that the function is energy-dependent leads to a 
modulation of the asymmetry $A$ in Equation~\ref{idealFivePar}.   
Finally, the rotation of the angular ($x'$) distribution modulates the 
phase $\phi_a$, because the average spin direction turns along with the
average momentum.   These additional modulations may be added to 
the fitting function of Equation~\ref{idealFivePar}, which now looks
substantially more complicated: 
\begin{eqnarray*}
f(t) & = & e^{-t/\tau} N \{[1 + C_1(t)] - A [1 + C_1(t) + C_2(t)]
  \cos(\omega_a t + \phi_a + C_3(t))\}  \\
C_{1,2,3}(t) &=&
 E(t) A_{c1,2,3} \cos(\omega_{CBO} t + \phi_{c1,2,3}) 
\end{eqnarray*}
$E(t)$ is an empirically-determined function that accounts for the the
coherence time of the CBO.
The factor $C_1(t)$ describes the overall modulation of the 
count rate, while $C_2(t)$ and $C_3(t)$ deal with the more subtle effects on
$A$ and $\phi_a$.

The systematic bias caused by these distortions of the time spectrum is
enhanced by an unfortunate coincidence of frequencies.  For the quadrupole
voltages used in the 1999 and 2000 running periods, 
$\omega_{CBO} \approx 465~{\rm kHz}$ is nearly $2\omega_a$.
Consequently, to the extent that the CBO introduce a background in the
time spectrum at the difference frequency $\omega_{CBO} - \omega_a$, they
perturb the fitted value of $\omega_a$. 
Figure~\ref{halfRing} shows the results of fits to the spectrum of each 
detector to a function that includes only the primary effect $C_1(t)$ of the 
CBO, with $C_2(t) = C_3(t) = 0$.  The fitted value of $\omega_a$ is not
consistent across detectors, but rather varies continuously by about 
$\pm 4$~ppm around the ring.  This behavior is explained by 
Figure~\ref{halfRingFFT}, which shows 
the Fourier transform of the difference between the time spectrum for the 
detectors on one half of the ring and the functional form to which it was fit.
A distinct peak at the sideband frequency $\omega_{CBO} - \omega_a$ is quite
visible when the spectrum is fit by the ideal function of 
Equation~\ref{idealFivePar}, and it is not eliminated by including only
$C_1(t)$.  With the full treatment including all three CBO effects,
this peak vanishes into the noise.  As the CBO bias is eliminated by fitting
for these effects, the apparent dependence of the value of $\omega_a$ 
on ring position also vanishes.

\section{Determination of $\omega_a$}

\begin{figure}[tb]
\begin{center}
\subfigure[]{\includegraphics[height=5.1cm]{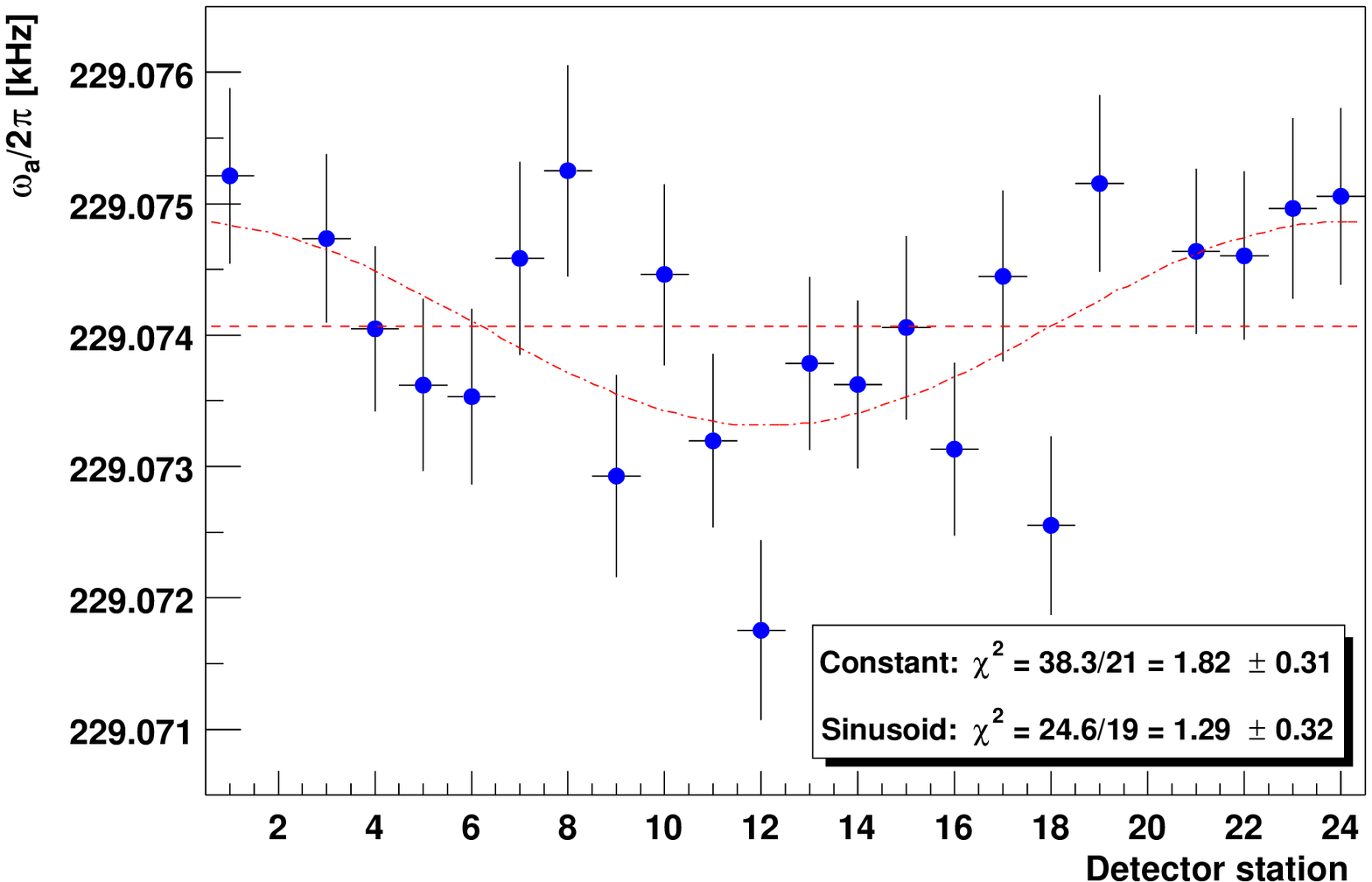} \label{halfRing}}
\subfigure[]{\includegraphics[height=5.1cm]{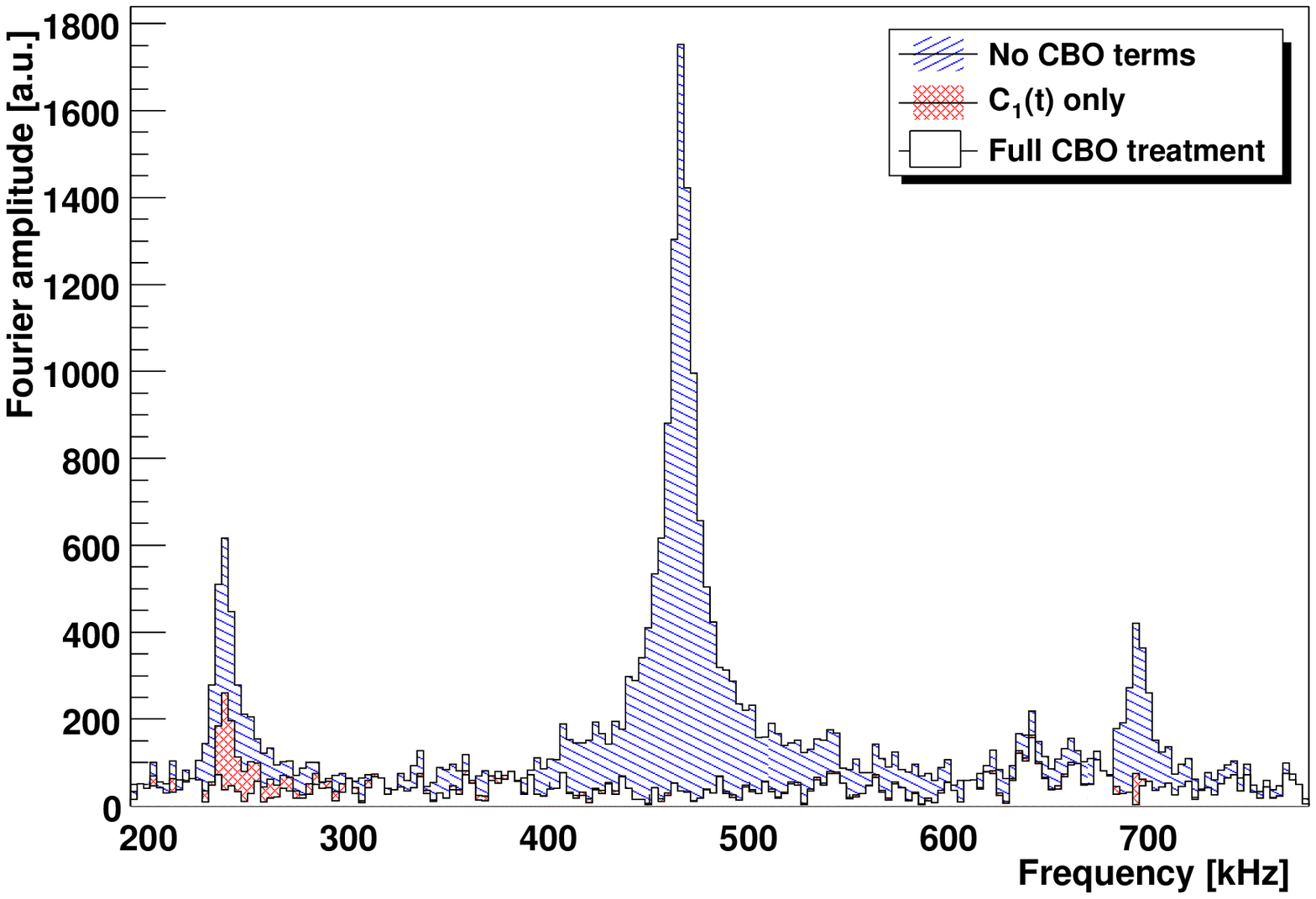} \label{halfRingFFT}}
\subfigure[]{\includegraphics[height=5.1cm]{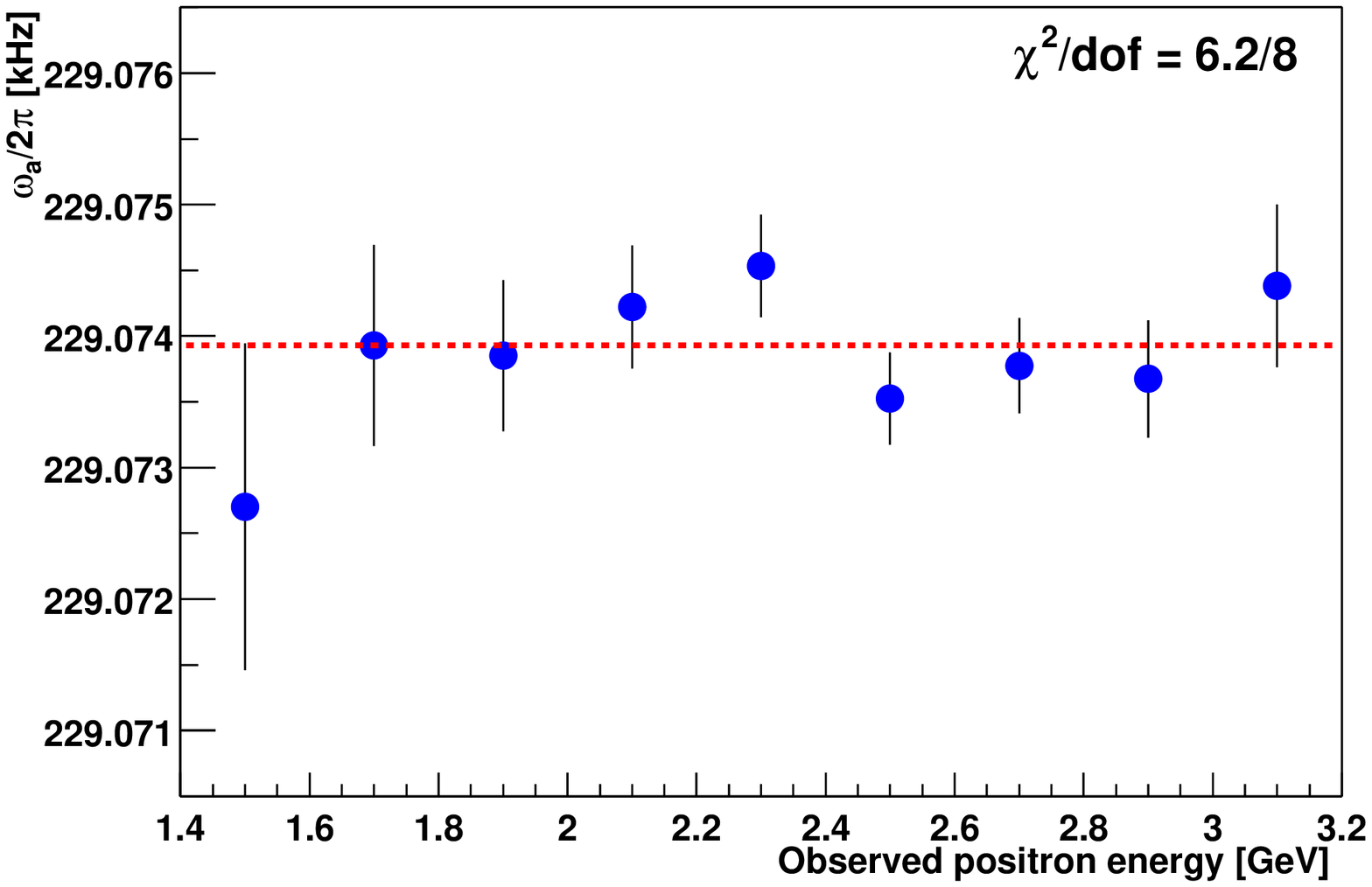}}
\subfigure[]{\includegraphics[height=5.1cm]{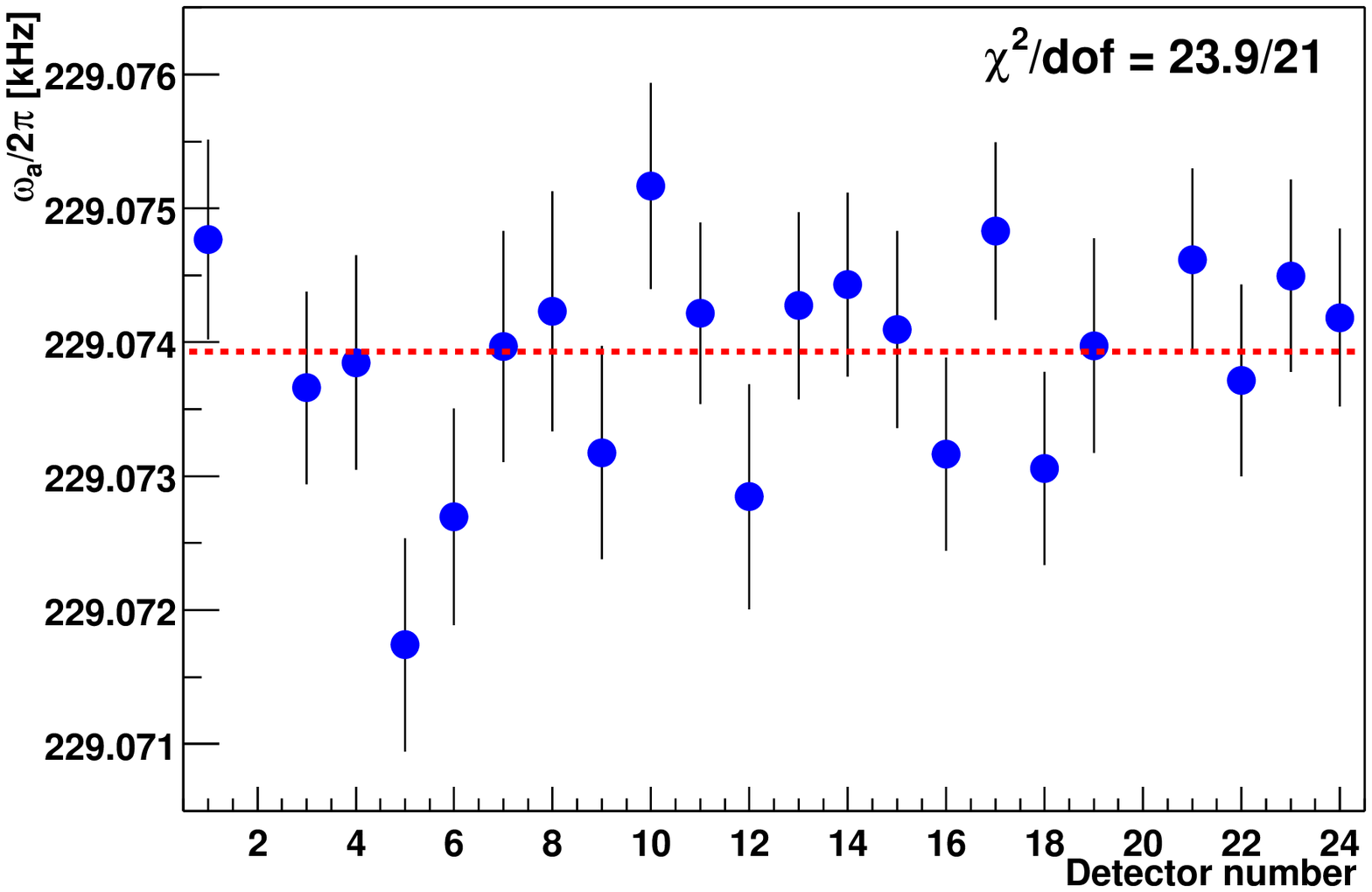}}
\end{center}
\caption{(a) Fitted value of $\omega_a$ as a function of detector station 
number with a simplified function where $C_2(t) = C_3(t) = 0$.  
(b) Fourier transform of the residuals following fits to the spectra of
detector stations 13-24.
(c) The fitted value of $\omega_a$ by energy bin and (d) by station,
indicating consistency in both cases.}
\label{omegaConsistency}
\end{figure}

Four independent extractions of $\omega_a$ were performed, based on two
independent implementations of the pulse fitting procedure.  The analyses
were done ``blind,'' without precise knowledge of the magnetic field.
Two of these analyses were conventional, setting a single energy threshold.  
One analysis formed a asymmetry signal by dividing the data into four subsets 
and combining them in a ratio that cancels out the exponential baseline.  
The fourth analysis divided the data into narrow energy bins, fit them
separately, and combined the results.  Some of the analyses included the
CBO modulations $C_2(t)$ and $C_3(t)$ in the fitting function, 
while others chose to rely on the reduction of the bias from the CBO by
an order of magnitude in the average of the detectors.  In the end, the 
results of the four analyses agreed with each other within statistical and 
systematic expectations.
The results of the energy-binned analysis, which included all CBO-related
terms, are shown in Figure~\ref{omegaConsistency}.  They demonstrate that the 
fitted value of $\omega_a$ is consistent in this case as a function of energy 
and detector number.  The four independent analyses were averaged to give
\begin{displaymath}
\omega_a/2\pi = 229\,074.11(0.14)(0.07)~{\rm Hz}
\end{displaymath}
This value has been corrected by 0.76~$\pm$~0.03~ppm to account for 
the effects of vertical oscillations and electric fields.
The significant contributions to the systematic uncertainty are, 
in addition to overlapping pulses and CBO, muon losses and detector gain 
variations during the measuring period.  The magnitudes of the contributions
from different sources varied to some extent among the four analyses, 
but all agreed on a total of 0.3~ppm. 

\section{Magnetic field measurement}

The magnetic field is mapped using NMR magnetometers,\cite{nmr} which measure
the spin precession frequency $\omega_p$ of protons at rest in the field.
A set of 17 of these devices is mounted on a trolley that is driven around the
inside of the storage region every few days during the running period,
yielding a multipole map as a function of azimuth.  Between trolley runs, 
changes in the field are tracked by fixed NMR probes located just outside
the vacuum chambers.  The trolley probes were calibrated against a standard
spherical water probe whose absolute calibration is known at the level of
0.05~ppm.\cite{Fei:1997}  
Over a period of several years, the field in the storage ring was iteratively 
shimmed for improved uniformity.  
Because the variation over the storage 
aperture is at the level 1~ppm, it is not necessary to know the distribution of 
muons very well.  

Two independent analyses of the $\omega_p$ data were conducted; again, they
were blind, performed without knowledge of $\omega_a$.  The analyses were
in agreement and found the result
\begin{displaymath}
\omega_p/2\pi = 61\,791\,595(15)~{\rm Hz}
\end{displaymath}
The largest contributions to the total error of 0.24~ppm are from 
the calibration of the trolley probes against the standard, 
non-linearity in the trolley position determination, and the reproducibility 
of the relationship between the field values measured by the trolley and
the fixed probes.

\section{Results and conclusions}

The ``world average'' value of $a_\mu$, which is dominated by the data set 
collected in 2000, is~\cite{Bennett:2002jb}
\begin{displaymath}
a_\mu^{exp;avg} = 11\,659\,203(8) \times 10^{-10} ~.
\end{displaymath}
Davier and collaborators provide two standard model theory results;
they differ in the experimental input used to the hadronic contributions.
They are~\cite{Davier:2002dy}
\begin{eqnarray*}
a_\mu^{th;ee} & = &  11\,659\,169.3(7.8) \times 10^{-10} ~{\rm and}~ 
a_\mu^{th;\tau} = 11\,659\,193.6(6.8) \times 10^{-10} ~.
\end{eqnarray*}
The first result gives a discrepancy of 3.0 standard deviations, while
the second indicates agreement at the level of 0.9 standard deviations.

At the moment, it does not seem appropriate to draw 
conclusions from the comparison of theory and experiment for $a_\mu$;
the theory value is still in flux.  The CMD-2 collaboration continues
to check its hadron production cross section results, and alternative
approaches such as radiative return~\cite{Venanzoni:2002jb} and lattice
calculations~\cite{Blum:2002ii} may hold some promise.  Also, Martin and Wells 
have demonstrated that it is possible to make some theoretical progress even 
with the current ambiguity.  They show~\cite{Martin:2002eu} that a significant 
part of the parameter space of the minimal supersymmetric standard 
model (MSSM) is excluded at the level of five standard deviations, even 
after assigning a very generous uncertainty to the hadronic effects.

\section*{Acknowledgments}

Funding for construction and operation of the experiment came from the
U.S. Department of Energy, the National Science Foundation,
the German {\em Bundesminister f{\"{u}}r Bildung und Forschung}, the Russian
Ministry of Science, and the U.S.-Japan Agreement in High Energy Physics.  
The computing facilities of the National Computational Science Alliance
were used for data analysis.
The author's tuition, stipend, and travel expenses were supported
by the National Science Foundation, by a GE Fellowship, and by a
Mavis Memorial Fund Scholarship award.

\section*{References}
\bibliography{moriond}

\end{document}